\documentclass[aps,twocolumn,pra,superscriptaddress,amsmath,amssymb]{revtex4-1}
\usepackage{dcolumn}
\usepackage{bm}
\usepackage{amsmath}
\usepackage{txfonts}
\usepackage[T1]{fontenc}
\usepackage{xspace}
\usepackage{ulem}
\usepackage{comment}
\usepackage{braket}
\ifx\pdfoutput\undefined
\usepackage[dvipdfmx]{graphicx}
\usepackage[dvipdfmx]{hyperref}
\usepackage[dvipdfmx]{color}
\else
\usepackage{graphicx}
\usepackage{hyperref}
\usepackage{color}
\fi

\begin{document}
\let\emph\textit

\title{
Hyperuniformity in two-dimensional periodic and quasiperiodic point patterns}

\author{Akihisa Koga}
\affiliation{
  Department of Physics, Tokyo Institute of Technology,
  Meguro, Tokyo 152-8551, Japan
}

\author{Shiro Sakai}
\affiliation{
  Center for Emergent Matter Science, RIKEN, Wako, Saitama 351-0198, Japan
}

\date{\today}
\begin{abstract}
  We study hyperuniform properties
  in various two-dimensional periodic and quasiperiodic point patterns.
  Using the histogram of the two-point distances, 
  we develop an efficient method to calculate the hyperuniformity order metric, 
  which quantifies the regularity of the hyperuniform point patterns.
  The results are compared with those calculated with 
  the conventional running average method.
  To discuss how the lattice symmetry affects the order metric,
  we treat the trellis and Shastry-Sutherland lattices
  with the same point density as examples of periodic lattices, 
  and Stampfli hexagonal and dodecagonal quasiperiodic tilings
  with the same point density
  as examples of quasiperiodic tilings.
  It is found that
  the order metric for the Shastry-Sutherland lattice
  (Stampfli dodecagonal tilings)
  is smaller than the other in the periodic (quasiperiodic) tiling,
  meaning that the order metric is deeply related to
  the lattice symmetry.
  Namely, the point pattern with higher symmetry is characterized
  by the smaller order metric when their point densities are identical.
  Order metrics for several other quasiperiodic tilings are also calculated.
\end{abstract}

\maketitle

\section{Introduction}
Quasiperiodic system has been the subject of extensive research since the discovery of the Al-Mn quasicrystal~\cite{Shechtman_1984}. 
The structures of the quasicrystals do not have the translational symmetry in the real space and are characterized 
by nontrivial rotation symmetries, 
which are forbidden in the conventional periodic lattices. 
Despite the aperiodicity, a quasiperiodic structure is completely ordered, 
leading to electron states distinct from periodic systems~\cite{Kohmoto_1986,Sutherland_1986,Kohmoto_1987,Arai_1988}.
However, aside from the rotational symmetry, it is not easy to capture the structural order of the quasiperiodic patterns.
Hyperuniformity is a framework to quantify the order of a point distribution
in a space~\cite{Torquato_2003,Torquato_2018} and is applicable to periodic,
quasiperiodic and random systems.
When the variance of the point density at a large length scale is smaller than a volume law,
the system is called hyperuniform, as described later in detail.
Hyperuniform point pattern is known to appear in nature, {\it eg.} 
the distribution of avian photoreceptors~\cite{Jiao_2014} and galaxy cluster~\cite{Philcox_2023}.
It has also been studied in applications such as photonic crystals~\cite{Florescu_2009,Florescu_2009_2,Man_2013},
which stimulates further investigations on the hyperuniform systems.
Periodic and quasiperiodic point distributions are known to be hyperuniform, 
and their order is, in most cases, characterized by a quantity called order metric. 
Then, the hyperuniformity order metrics 
for some periodic and quasiperiodic point patterns
have been studied~\cite{Torquato_2003,Zachary_2009,Lin_2017,Oguz_2017,Torquato_2018}.
Recently, electronic properties on a quasiperiodic structure have been discussed 
in terms of hyperuniformity~\cite{Sakai_Arita_Ohtsuki_2022,Sakai_2022}.
The hyperuniformity may also be useful for characterizing the spatial distribution of the order parameter in broken-symmetry phases~\cite{Jagannathan_1997,Wessel_2003,Koga_2017,Sakai_Takemori_Koga_Arita_2017,Araujo_2019,Sakai_Arita_2019,Cao_2020,Inayoshi_2020,Takemori_2020,Koga_2020,Ghadimi_2021,Koga_2021,Sakai_Koga_2021,Inayoshi_2022,Koga_Coates_2022}. 
Therefore, it is instructive to give the order metrics 
for several periodic and quasiperiodic point patterns as references.

In this paper, we study the hyperuniformity  
in several two-dimensional periodic and quasiperiodic lattices.
We develop an efficient method to calculate the order metric quantitatively,
exploiting the histogram of two-point distances and the filter function. 
We apply the method to the periodic and quasiperiodic lattices composed of the squares and triangles 
such as Shastry-Sutherland, trellis,
and hexagonal and dodecagonal Stampfli lattices~\cite{Stampfli}.
We clarify that the point pattern with the higher rotational symmetry
has a smaller order metric when the point density is identical. 
Furthermore, we study the effect of the depletion in the point pattern.
Then, we find that the larger scale calculations are necessary 
to precisely obtain the order metric of the depleted lattices.
This should be crucial to obtain the order metrics for some quasiperiodic point patterns.
We provide precisely calculated values of the order metrics for various periodic and quasiperiodic
point patterns.

This paper is organized as follows.
In Sec.~\ref{Hyper},
we briefly explain the hyperuniformity and
define the order metric to characterize the regularity of the lattices.
We also explain the detail of our methods to efficiently obtain the order metric.
We demonstrate the benchmark of our method in Sec.~\ref{benchmark}.
Numerical results for various periodic and quasiperiodic lattices are shown in Sec.~\ref{Result}.
A summary is given in the last section.

\section{Hyperuniformity order metric}\label{Hyper}

In this study, we focus on the hyperuniformity to characterize point patterns
in two dimensions~\cite{Torquato_2003}.
First, we consider the circular window $\Omega$
in the two-dimensional Euclidean space whose center and radius are denoted as ${\bf X}$ and $R$,
respectively.
The number of points inside the domain $N_{\bf X}(R)$ depends on the coordinate ${\bf X}$.
When one examines it for sufficiently many coordinates ${\bf X}$,
its average $\langle N_{\bf X}(R)\rangle$ is proportional to the density of points $\rho$,
as $\langle N_{\bf X}(R)\rangle=\pi \rho (R/a)^2$,
where $a$ is the length scale of the point pattern
({\it e.g.} the distance between neighboring points).
The corresponding variance,
\begin{align}
  V(R)&=\langle N_{\bf X}^2(R)\rangle-\langle N_{\bf X}(R)\rangle^2,\label{eq:bare}
\end{align}
on the other hand, strongly reflects the spatial structure of the point pattern.
In the large $R$ limit,
the variance may be represented as 
\begin{align}
  V(R)=AR^2+BR+\cdots,
\end{align}
with coefficients $A$ and $B$.
It is known that $A$ is finite in the case of
the randomly distributed point pattern~\cite{Torquato_2003}.
By contrast, for all periodic and most of quasiperiodic point patterns,
$A=0$ and $B\neq 0$.
In this case, the system is called hyperuniform (class I)
and the coefficient $B$ characteristic of the point pattern
is called an order metric.
Here, we systematically examine the order metrics for various periodic and quasiperiodic lattices.

To evaluate the order metric, 
we define the function $\Lambda(R)$ as
\begin{align}
  \Lambda(R)=\frac{V(R)}{R},
\end{align}
for two-dimensional hyperuniform point patterns and
\begin{align}
  B = \lim_{R \rightarrow\infty}\Lambda(R).
\end{align}
According to the equation (62) in Ref.~\cite{Torquato_2003},
$\Lambda(R)$ is given as 
\begin{align}
\Lambda(R)=4\phi \left(\frac{R}{a}\right)\left[1-4\phi \left(\frac{R}{a}\right)^2+\frac{1}{N}\sum_{i\neq j}\alpha(r_{ij};R)\right], 
\label{eq:7}
\end{align}
with $\phi=\pi\rho/4$, $r_{ij}(=|{\bf r}_i-{\bf r}_j|)$ is the distance between $i$th and $j$th points, and
\begin{align}
  \alpha(r;R)&=\left\{
  \begin{array}{ll}
    \displaystyle\frac{2}{\pi}\left[\cos^{-1}\frac{r}{2R}-\frac{r}{2R}\left(1-\frac{r^2}{4R^2}\right)\right] & (r\le 2R)\\
    \displaystyle 0 & (r>2R)
  \end{array}
  \right..
\end{align}
Note that $\phi$ corresponds to the packing factor $F$ 
when the minimal two-point distance equals $a$.
The sum in Eq. (\ref{eq:7}) is taken for all two points in the point pattern.
This means that one can evaluate $\Lambda(R)$ 
without performing the samplings of the circular windows in the two dimensional space.
We note that $\bar{\Lambda}(R)=\Lambda(R)\phi^{-1/2}$ 
is the scale-independent function, which will be mainly discussed in the following.

In the ordered point pattern, the two-point distance $r_{ij}$ takes certain discrete values.
This allows us to introduce its histogram as
\begin{align}
  h(r)&=\frac{1}{N}\sum_{i\neq j}\delta_{r,r_{ij}}=\sum_k \omega_k \delta_{r,r_k},
\end{align}
where $\omega_k$ is the weight of the two-point distance $r_k$.
By means of the histogram, we obtain
\begin{align}
\bar{\Lambda}(R)&=2\sqrt{\pi\rho} \left(\frac{R}{a}\right) \left[1-\pi\rho \left(\frac{R}{a}\right)^2+\sum_k \omega_k\alpha(r_k;R)\right],\\
\psi(R) &=\frac{1}{\pi}\left(\frac{a}{R}\right)^2\left[1+\sum_k\omega_k\theta(R-r_k)\right],
\end{align}
where $\psi(R)$ is the average point density in the circular region
centered at a point with radius $R$.
We note that, for the periodic lattices, we evaluate the histogram $h(r)$ with a set of $\{r_k,\omega_k\}$,
focusing on each inequivalent point in the unit cell and
calculating the distances between it and other points in the entire space.

The periodic and quasiperiodic point patterns have the length scale due to its regularity,
leading to oscillation behavior in $\bar{\Lambda}(R)$ and $\psi(R)$ in the scale of $a$.
Nevertheless, $\psi(R)$ well converges to $\rho$ in the limit $R\rightarrow\infty$.
This may be useful to confirm the precision of the histogram.
By contrast, $\bar{\Lambda}(R)$ always oscillates with respect to $R$.
Therefore, the running (cumulative moving) average~\cite{Kim_2017}
may be useful to deduce its average, as
\begin{align}
  f_1(R)=\frac{1}{R-R_0}\int_{R_0}^R f(r)dr,
\end{align}
where $f_1(R)$ is the running average and $f=\bar{\Lambda}$ or $\psi$.
The constant $R_0$ is set to zero in this study.
Now, we propose another way to evaluate the average as
\begin{align}
f_2(R)=\frac{\displaystyle\int_{0}^\infty f(r) g(r;R)dr}{\displaystyle\int_{0}^\infty g(r;R)dr},\label{f2}
\end{align}
where $g(r;R)$ is the filter function. 
If $g(r; R)=\theta(R-r)$, $f_2(R)$ is reduced to $f_1(R)$.
Practically, we use the Gauss function as a filter function, {\it i.e.}
\begin{align}
    g(r;R)=\frac{1}{\sqrt{\pi}\sigma}\exp\left[-\left(\frac{r-R}{\sigma}\right)^2\right],
\end{align}
where $R$ is the center of the Gaussian and $\sigma$ is its width.

For a finite system size tractable with a numerical calculation, 
the running average $f_1$ is appropriate to roughly examine the order metric,
but strongly depends on the endpoints of the integral
since $\bar{\Lambda}$ and $\psi$ always oscillate with respect to $R$.  
This also yields oscillation behavior in $f_1$, and
it becomes difficult to obtain the order metric precisely.
On the other hand, the Gaussian function in the filter method
strongly suppresses the oscillations at the endpoints of the integral,
which allows us to evaluate the order metric precisely.
Furthermore, $f_2(R)$ with a finite $R$ little depends on the local structure $(R\sim 0)$
since it is evaluated for the integral around $R$, $(R-\sigma, R+\sigma)$,
in contrast to the running average $f_1$. 
Therefore,
this filter function method is expected to suppress oscillations 
in $\bar{\Lambda}(R)$ and $\psi(R)$,
giving precise values of the averages at reasonable numerical costs,
as we shall demonstrate in the following.

\section{Benchmark of the method}\label{benchmark}
Here, we demonstrate the benchmark of our method.
\begin{figure}[htb]
  \includegraphics[width=\linewidth]{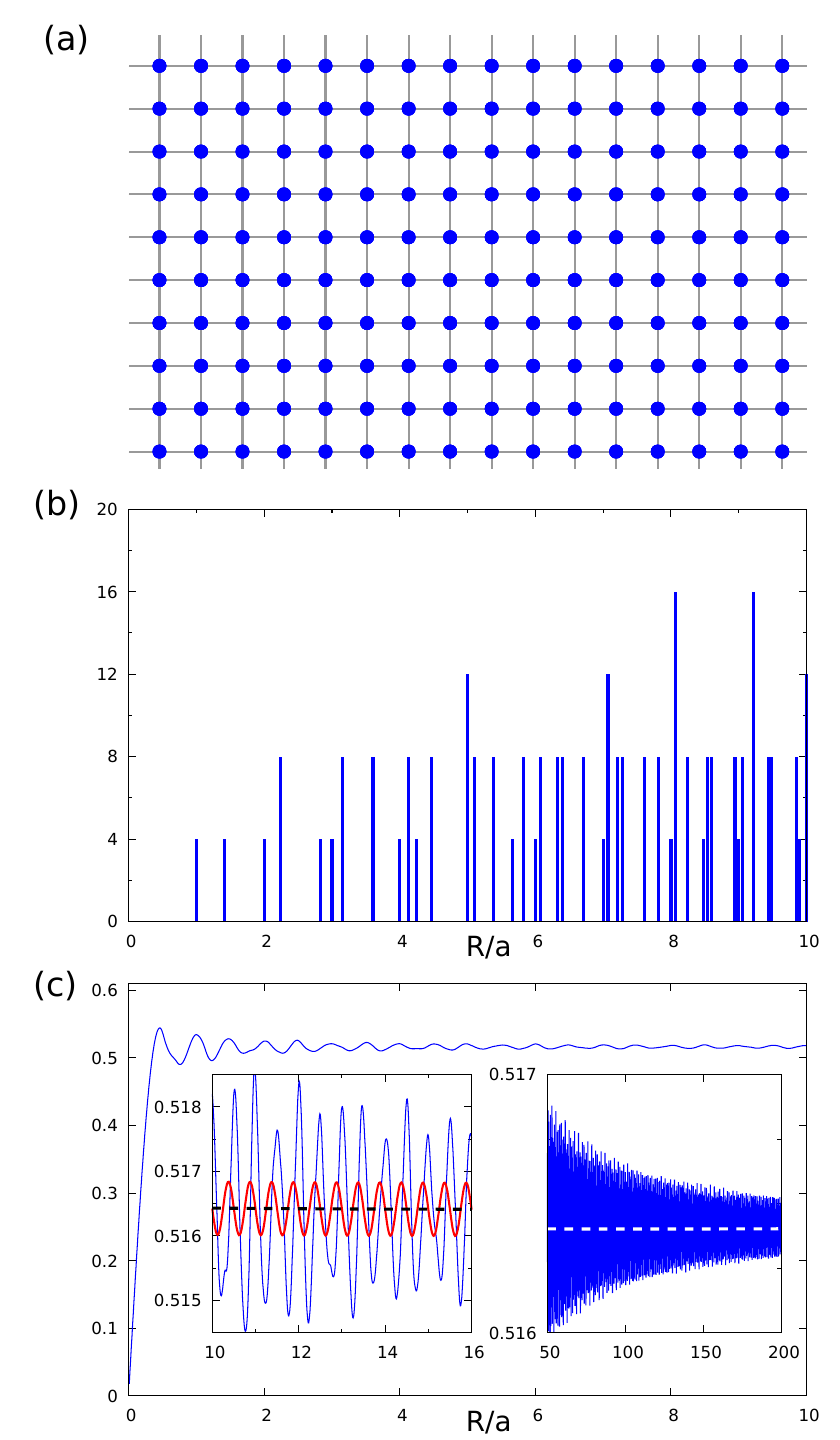}
  \caption{
    (a) Square lattice and (b) its histogram $h(R)$.
    (c) Running average $\bar{\Lambda}_1(R)$
    for the square lattice.
    Left inset: Thin blue curve shows the result of the running average 
    while red bold solid and black dashed curves 
    indicate $\bar{\Lambda}_2(R)$ obtained by the Gaussians
    with $\sigma/a=0.4$ and $\sigma/a=2.0$, respectively.
    Right inset: Blue solid curve shows the running average while white dashed line shows the result with $\sigma/a=2.0$.
    }
  \label{fig:square}
\end{figure}
As a simple example, we start with the square lattice with a lattice constant $a$, 
which is shown in Fig.~\ref{fig:square}(a).
Figure~\ref{fig:square}(b) shows the histogram as a function of the distance $R$.
We see that $\omega_k$ takes only multiple of four, reflecting the fourfold rotational symmetry.
The running average $\bar{\Lambda}_1(R)$
is shown in Fig.~\ref{fig:square}(c).
It is seen that the running average approaches a certain value for $R/a\gtrsim 2$,
while oscillations remain up to a much larger $R$, as shown in the insets.
Therefore, careful treatments are necessary to obtain the order metric precisely.
We also show the results $\bar{\Lambda}_2(R)$ obtained
by means of the Gaussian with $\sigma/a=0.4$ and $2.0$
as the bold solid and dashed curves in the left inset of Fig.~\ref{fig:square}(c).
Small oscillations appear for $\sigma/a=0.4$, while
it is negligible for $\sigma/a=2.0$.
The latter result seems almost constant up to $R/a=200$, which is shown in the right inset.
This contrasts with larger oscillation behavior observed 
in the running average $\bar{\Lambda}_1(R)$.
The results for several choices of Gaussian parameters are 
shown in Tab.~\ref{tab}.
We find that
the obtained values are well converged to five digits around $R/a\sim 100$ and 
$\sigma/a \sim 5$.
The normalized order metric $\bar{B}=B\phi^{-1/2}$ is obtained as
$\bar{B}\sim 0.51640$, which is in good agreement with $0.516401$ 
obtained in the pioneer work~\cite{Torquato_2003}.
We also note that, even with a smaller $\sigma$, reasonable results
are obtained, by taking into account the width of the oscillations in $\bar{\Lambda}_2(R)$.
\begin{center}
  \begin{table}[htb]
    \caption{The normalized order metric $\bar{\Lambda}_2$ and density of points $\psi_2$
    for the square lattice
      obtained by means of the Gaussian filters $g(r;R)$ with $R$ and $\sigma$.
     }
    \centering
    {
    \renewcommand\arraystretch{1.2}
    \begin{tabular}{cccc || cccc}
      \toprule
      $R/a$ & $\sigma/a$ & $\bar{\Lambda}_2 $ & $\psi_2$ & $R/a$ & $\sigma/a$ & $\bar{\Lambda}_2 $ & $\psi_2$ \\
      \hline
      5 & 0.2 &  0.512 661 & 0.976 985 & 30 & 1.0 & 0.516 404 & 1.000 000\\
      10 & 0.2 & 0.512 530 & 0.994 516 & 30 & 5.0 & 0.516 404 & 1.000 000\\
      10 & 0.5 & 0.516 429 & 0.999 149 & 50 & 5.0 & 0.516 402 & 1.000 000\\
      10 & 1.0 & 0.516 429 & 1.000 000 &100 & 5.0 & 0.516 402 & 1.000 000\\
      \toprule
    \end{tabular}
    \label{tab}
    }
  \end{table}
\end{center}

The quasiperiodic point pattern 
has unique properties
distinct from the periodic and disordered patterns.
One of them is the repeated structure in the tiling, which
is known as the Conway's theorem for the Penrose tiling~\cite{Bruijn,Gardner}.
In generic quasiperiodic patterns, any finite part of the point pattern repeats itself 
within a finite distance proportional to its diameter.
Therefore, in the thermodynamic limit, the circular region with a radius $R$
appears ubiquitously at a density $\sim O(R^{-2})$.
For a circular window of radius $R$ centered at a point, 
there are only a finite number of the possible point patterns inside the window, 
while the number increases with increasing $R$.
Since it is hard to evaluate analytically $\{ r_k, \omega_k\}$
for the large $R$,
we here deduce the histogram by means of random sampling.

We briefly explain the detail of our sampling.
As discussed above, a large number of samples are necessary
to deduce the order metric for the quasiperiodic tilings.
To this end, we use the inflation-deflation rule
to systematically generate the point patterns
around an arbitrary coordinate.
In each sampling, we first choose a coordinate ${\bf X}$
in the squared area $L\times L$. 
Then, we randomly choose a point ${\bf x}_i$
in the circular region centered at ${\bf X}$ with a large radius $R_{samp}$.
Finally, we obtain the vertices ${\bf x}_{j}$ in the circular window
centered at ${\bf x}_i$ and
calculate the set $\{r_j, \omega_j\}$ with $r_j=|{\bf x}_i-{\bf x}_{j}|$.
Sampling many times $N_{samp}$, 
we obtain the histogram $h(r)$.
In the following, we set $L=10^8a$, $R_{samp}/a=4000$, and $N_{samp}>10^9$.

\begin{center}
  \begin{table}[htb]
    \caption{The normalized order metric $\bar{\Lambda}_2$ and the point density $\psi_2$ 
    of the Penrose tiling,
    obtained by means of the Gaussian filters $g(r;R)$ with the listed $R/a$ and $\sigma/a$. }
    \centering
    {
    \renewcommand\arraystretch{1.2}
    \begin{tabular}{cccc || cccc}
      \toprule
      $R/a$ & $\sigma/a$ & $\bar{\Lambda}_2 $ & $\psi_2$ & $R/a$ & $\sigma/a$ & $\bar{\Lambda}_2$ & $\psi_2$ \\
      \hline
      5 & 0.2 &  0.579 62 & 1.234 05 & 30 & 1.0 & 0.591 56 & 1.231 07 \\
      10 & 0.2 & 0.605 01 & 1.235 54 & 30 & 5.0 & 0.591 45 & 1.231 07 \\
      10 & 0.5 & 0.593 41 & 1.231 66 & 50 & 5.0 & 0.591 45 & 1.231 07 \\
      10 & 1.0 & 0.591 62 & 1.231 09 &100 & 5.0 & 0.591 44 & 1.231 07 \\
      \toprule
    \end{tabular}
    \label{tab2}
    }
  \end{table}
\end{center}

\begin{figure}[htb]
  \includegraphics[width=\linewidth]{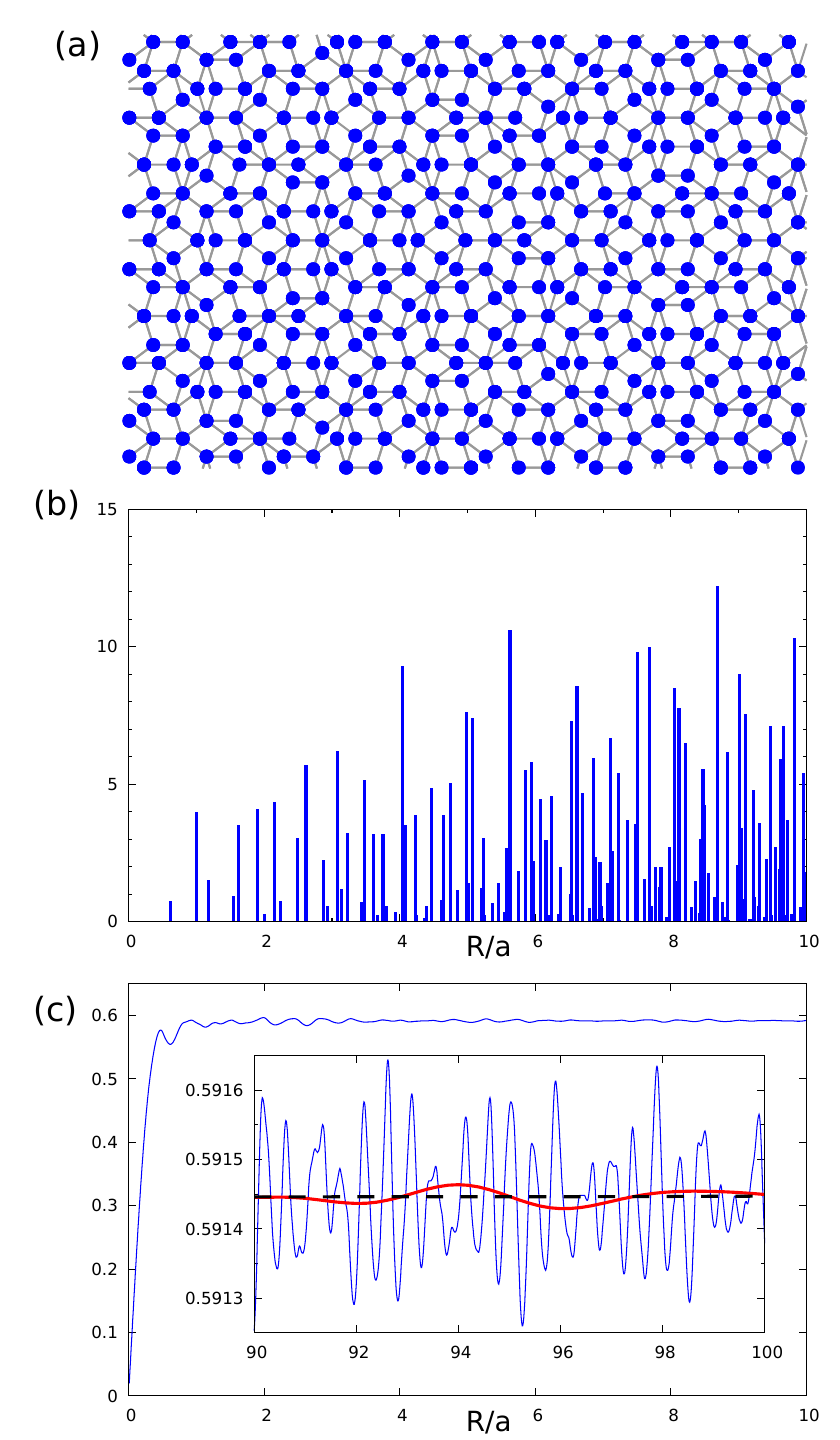}
  \caption{
    (a) Penrose tiling and (b) its histogram $h(R)$.   
    (c) The normalized running averages $\bar{\Lambda}_1(R)$
    for the Penrose tiling. 
    Red bold solid and black dashed lines in the inset represent 
    the results $\bar{\Lambda}_2(R)$
    obtained from the Gaussian filters with $\sigma/a=2$ and $\sigma/a=5$. 
  }
  \label{fig:penrose}
\end{figure}

Here, we demonstrate the results for the Penrose tiling
as an example of the quasiperiodic tilings,
which is shown in Fig.~\ref{fig:penrose}(a).
This tiling is composed of the skinny and fat rhombuses with edge length $a$.
Figure~\ref{fig:penrose}(b) shows the histogram of Penrose tiling.
This is in a good agreement with the analytical results for small $R$,
$(R_k/a, \omega_k)=(1/\tau, 2/\tau^2), (1,4), (\sqrt{3-\tau},4/\tau^2),
(\sqrt{3-1/\tau}, 4/\tau^3), (\tau, 2+4/\tau^2), (\sqrt{\tau+2}, 4+2/\tau^6), (2, 2/\tau^4)$
with the golden ratio $\tau[=(1+\sqrt{5})/2]$.
We find that 
the number of peaks in the histogram is much larger than that for the square lattice,
as discussed before.
The running average approaches a certain value around $R/a\sim 2$, 
while oscillates around $\bar{\Lambda}_1(R)\sim 0.6$,
as shown in Fig.~\ref{fig:penrose}(c).
The results obtained by means of the Gaussian filters with $\sigma/a=2$ and $5$ are
shown as the solid and dashed curves in the inset.
Invisible oscillations appear in the results with $\sigma/a=5$.
The numerical results for some sets of $\{R, \sigma\}$ are shown in Tab.~\ref{tab2}.
We find that
the obtained values are well converged to five digits around $R/a\sim 100$ and $\sigma/a \sim 5$,
which is similar to the case of the square lattice.
We obtain the order metric $\bar{B}\sim 0.59145$. 
This is smaller than the result $\bar{B}=0.60052$ in Ref.~\cite{Zachary_2009},
but is consistent with the recent result~\cite{Lin_2017}.

We wish to note that the histogram with $R/a\lesssim 10$
leads to a reasonable value of the order metric to three digits,
as shown in Tabs.~\ref{tab} and \ref{tab2}.
This fact would be useful in evaluating the order metric
for small-size point patterns observed in the experiments. 

\section{Numerical Results}\label{Result}
\subsection{Periodic point patterns}

\begin{center}
  \begin{table}[htb]
    \caption{Densities of points $\rho$, $\phi(=\pi\rho/4)$, and normalized order metrics $\bar{B}$ 
    for various periodic point patterns. 
    }
    \label{table3}
    \centering
    {
    \renewcommand\arraystretch{1.2}
    \begin{tabular}{l|ll|l|l}
        \toprule
        Pattern & $\rho $ && $\phi$ & $\bar{B}$ \\
      \hline
        Triangular & $2/\sqrt{3}$ & 1.154 701& 0.906 900 & 0.508 35\\
        Square & 1 & 1.000 000 & 0.785 398 & 0.516 40 \\
        Honeycomb & $4/(3\sqrt{3})$ & 0.769 800 &  0.604 600 & 0.566 99\\
        Kagome & $\sqrt{3}/2$ & 0.866 025 & 0.680 175 & 0.586 99 \\
        1/5-depleted square & 4/5 & 0.800 000 & 0.628 319 & 0.604 62 \\
        Shastry-Sutherland & $8-4\sqrt{3}$ & 1.071 797 & 0.841 787  & 0.516 64\\
        Trellis & $8-4\sqrt{3}$ & 1.071 797 & 0.841 787 &  0.518 77\\
        \toprule
    \end{tabular}
    }
  \end{table}
  \end{center}

We first demonstrate the order metrics for
the square, triangular, honeycomb, and Kagome lattices
as simple periodic lattices.
Combining the histogram and filter function methods,
we obtain the results shown in Table~\ref{table3}.
Here, we have set $a$ as a distance between the nearest neighbor point pairs.
These are in a good agreement with those
in the pioneer works~\cite{Torquato_2003,Zachary_2009},
except for that for the honeycomb lattice.

\begin{figure}[htb]
  \includegraphics[width=\linewidth]{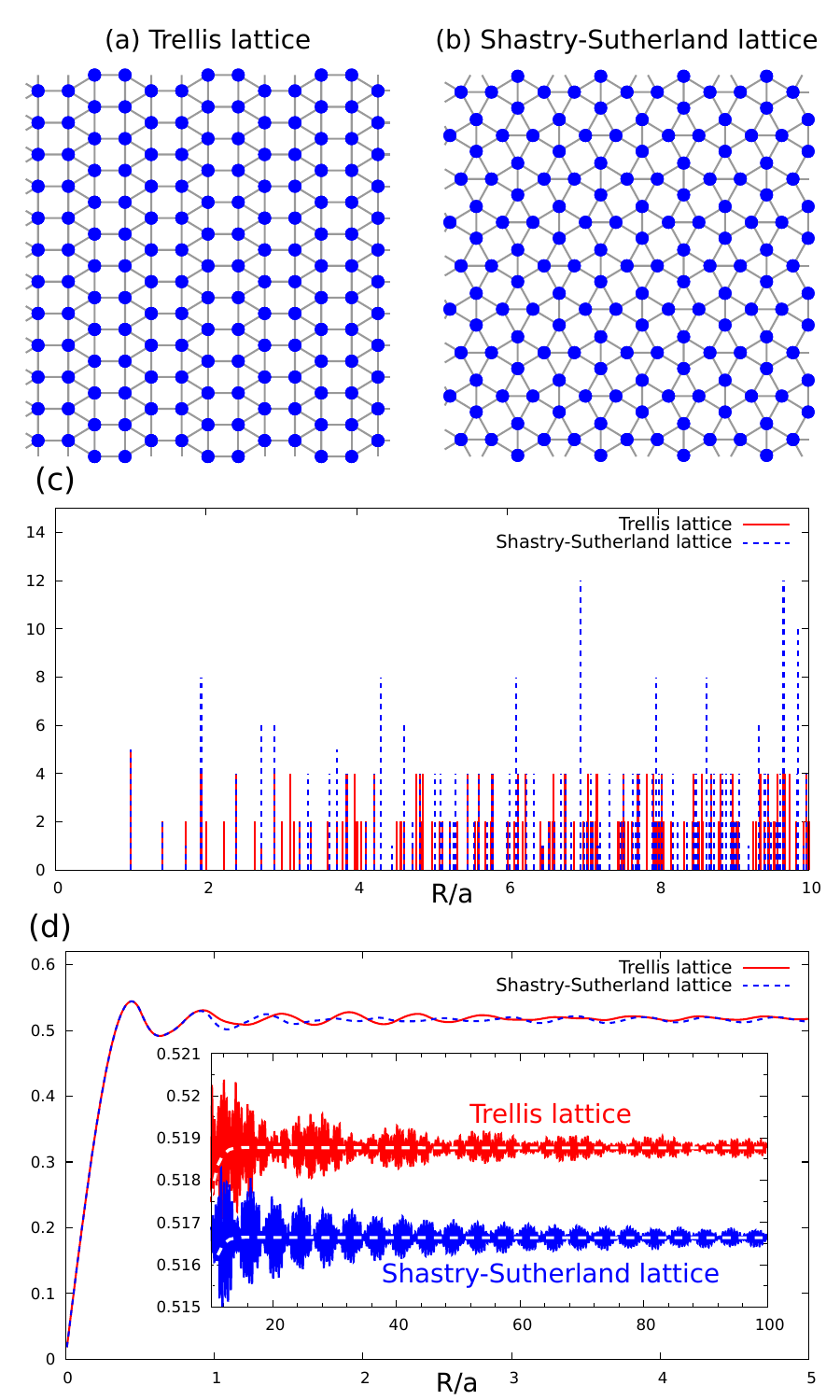}
  \caption{
    (a) Trellis and (b) Shastry-Sutherland lattices,
    and (c) their histograms.
    (d) Solid (dashed) lines represent the running average $\bar{\Lambda}_1(R)$
    for the trellis (Shastry-Sutherland) lattice.
    In the inset, we compare the running averages $\bar{\Lambda}_1(R)$ (solid lines)
    and $\bar{\Lambda}_2(R)$ (dashed lines) with $\sigma/a=5$.
  }
  \label{fig:treSS}
\end{figure}

Now, we discuss how the order metric is affected by the rotational symmetry.
To this end, we consider the trellis and Shastry-Sutherland lattices,
which are both composed of the triangles and squares with the edge length $a$.
The latter is also known as the $\sigma$ phase in the metallurgy.
The lattice structures are schematically shown in Figs.~\ref{fig:treSS}(a)
and \ref{fig:treSS}(b).
The quantum spin systems on these lattices are known as
geometrically frustrated systems, and 
theoretical and experimental studies have been done 
so far~\cite{exLadder1,exLadder2,exSS,Gopalan,ShastrySutherland,MiyaharaUeda,Koga},
where distinct magnetic properties are discussed.
From the structual point of view,
in both lattices, the volumes of Voronoi cells $V$,
atomic packing factor $F$, densities of triangles and squares are identical: 
$V/a^2=1/\rho=(2+\sqrt{3})/4$, $f=\pi/(2+\sqrt{3})$,
$\rho_\triangle=2/3$, and $\rho_\square=1/3$.
This allows us to discuss how the symmetry of the point pattern
affects the order metric.
Namely, the point pattern of the trellis lattice belongs to 
$D_2$ point group, while that of the Shastry-Sutherland lattice belongs to 
$C_4$ point group.

As shown in Figs.~\ref{fig:treSS}(a) and \ref{fig:treSS}(b), 
each point is shared by two squares and three triangles.
Therefore, no difference appears in the histogram when $R/a<\sqrt{3}$:
$(R/a, \omega)=(1,5)$ and $(\sqrt{2},2)$, as shown in Fig.~\ref{fig:treSS}(c).
On the other hand, the difference appears in
the coordination number for the third nearest-neighbor
with a distance $R/a=\sqrt{3}$.
Namely, $\omega=1$ for the Shastry-Sutherland lattice,
while $\omega=2$ for the trellis lattice.
Beyond $R/a=\sqrt{3}$, a finite weight appears at a fewer values of $R$
in the Shastry-Sutherland lattice,
which reflects the higher rotational symmetry.
Then, $\bar{\Lambda}(R)$ and its running average for both lattices
are identical for $R/a<\sqrt{3}/2$ and 
move apart beyond it, which are shown in Fig.~\ref{fig:treSS}(c).
Finally, we clearly find that $\bar{\Lambda}_1(R)$ takes distinct values around $R/a=100$.
The results obtained by the filter function method with $\sigma/a=5$ are shown 
as the dashed lines in the inset of Fig.~\ref{fig:treSS}(c).
The invisible oscillations appear and 
we obtain $\bar{B}=0.51664$ for the Shastry-Sutherland lattice and
$\bar{B}=0.51877$ for the trellis lattice (see Tab.~\ref{table3}).
This result may be explained by the difference of the rotational symmetry of the point patterns.
Namely, the rotational symmetry for the Shastry-Sutherland lattice
is higher than the other,
which results in the smaller order metric
{\it i.e.} higher regularity.

\begin{figure}[htb]
  \includegraphics*[width=\linewidth]{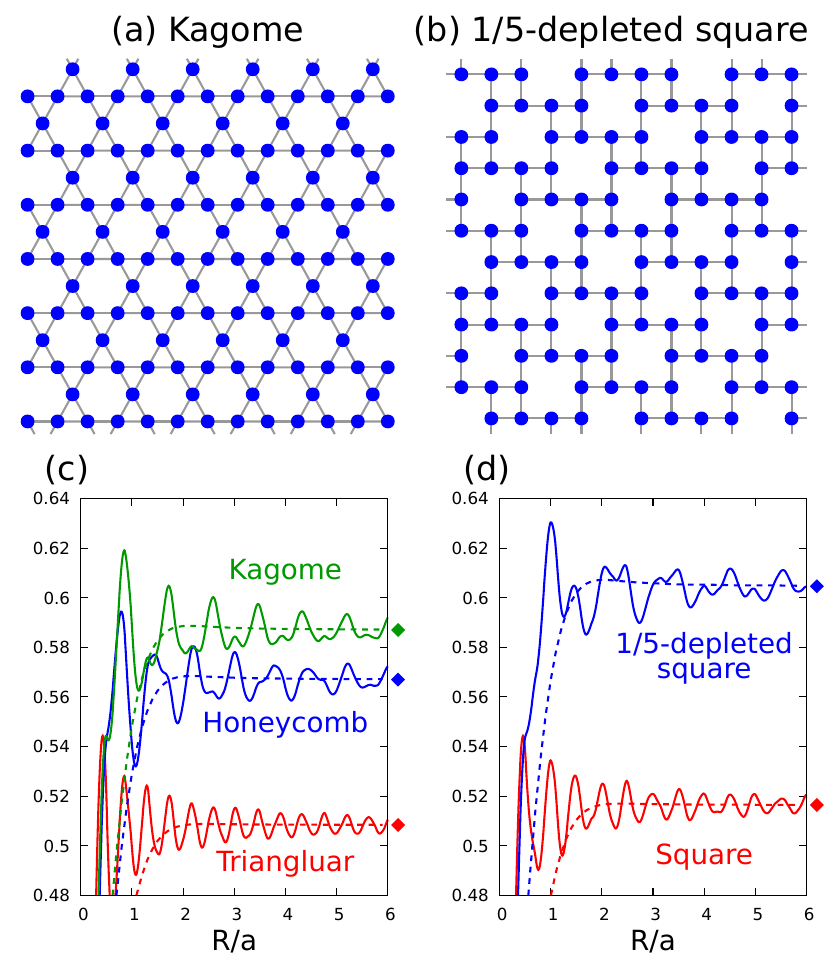}
  \caption{
    (a) Kagome and (b) 1/5-depleted square lattices.
    (c) [(d)]
    Solid lines represent the running averages $\bar{\Lambda}_1(R)$ 
    for the Kagome, honeycomb and triangular lattices
    (1/5-depleted square and square lattices) for $R/a\le 6$.
    Dashes lines represent $\bar{\Lambda}_2(R)$ with $\sigma/a=1.0$
    and diamonds represent the normalized order metrics $\bar{B}$ shown in Tab.~\ref{table3}.
  }
  \label{fig:defects}
\end{figure}
We also discuss how the order metric is affected by the depletion in the lattice.
To this end, we deal with the triangular and square lattices,
and consider their depleted lattices.
The honeycomb and Kagome lattices can be regarded as the 1/3- and
1/4-depleted triangular lattices, respectively.
The Kagome and 1/5-depleted square lattices
are schematically shown in Figs.~\ref{fig:defects}(a) and \ref{fig:defects}(b), respectively.
The running averages for $R/a<6$ are shown in Figs.~\ref{fig:defects}(c)
and \ref{fig:defects}(d).
We find that the running averages for triangular and square lattices
increase rapidly for $R/a\lesssim 0.5$ and 
tend to converge to certain values with decaying oscillation.
By contrast, the running averages for the depleted lattices
tend to slowly increase with oscillation even for a larger $R$. 
The amplitude of the oscillation looks larger for the depleted lattices.
Thus, introducing the depletion (space) into the lattice,
the running average slowly increases with $R$, and 
it becomes more difficult to precisely evaluate the order metric.
This should be important for the several quasiperiodic tilings
since they are composed of multiple tiles with distinct areas.
By means of the filter functions with $\sigma/a=1.0$, 
we obtain the smooth curves shown as the dashed lines
in Figs.~\ref{fig:defects}(c) and \ref{fig:defects}(d).
The order metrics for the above depleted lattices are shown as the diamonds 
and in Tab.~\ref{table3}.

\subsection{Quasiperiodic point patterns}
We next consider several quasiperiodic point patterns.
\begin{figure}[htb]
  \includegraphics*[width=\linewidth]{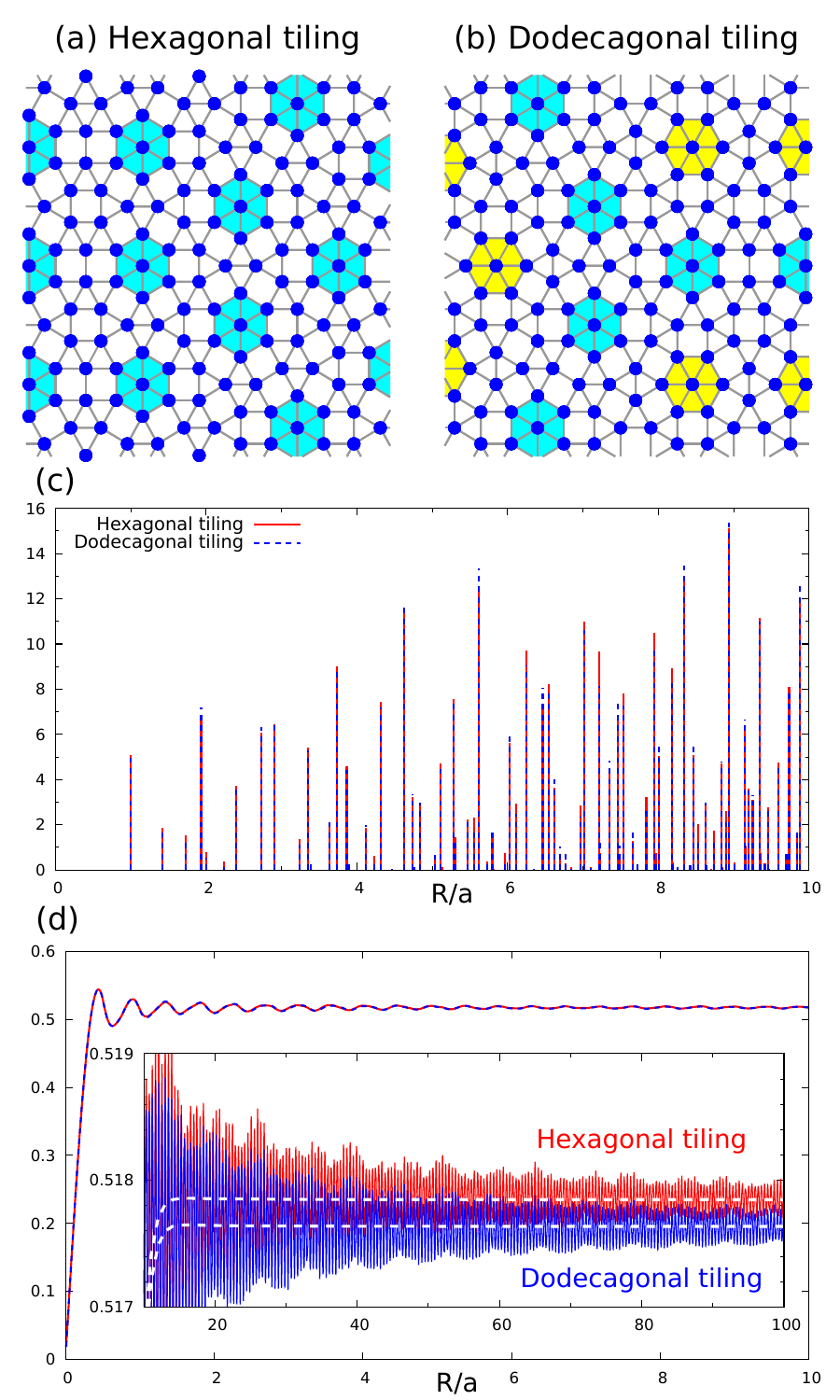}
  \caption{
    (a) Stampfli hexagonal and (b) dodecagonal tilings,
    and (c) their histograms.
    (d) Solid (dashed) lines represent the running average $\bar{\Lambda}_1(R)$
    for the Stampfli hexagonal (dodecagonal) tiling. 
    In the inset, we compare the running averages $\bar{\Lambda}_1(R)$ (solid lines) 
    and $\bar{\Lambda}_2(R)$ (dashed lines) with $\sigma/a=5$.
  }
  \label{fig:Stampfli}
\end{figure}
First, we deal with the Stampfli hexagonal and dodecagonal tilings~\cite{Stampfli}
composed of the triangles and squares with the edge length $a$,
which are shown in Figs.~\ref{fig:Stampfli}(a) and \ref{fig:Stampfli}(b).
These tilings are similar to each other, but a difference
appears in the hexagonal structure composed of six adjacent triangles,
which are shown as shaded areas in Figs.~\ref{fig:Stampfli}(a) and \ref{fig:Stampfli}(b).
Namely, two edges of each hexagon are always parallel to $y$-axis
for the Stampfli hexagonal tilings, 
but the other equally includes two hexagonal structures with distinct directions.
This difference hardly affects local properties,
but leads to the difference in the global rotational symmetry.
In fact, no difference appears in the histogram when $R/a<\sqrt{3}$:
$(R/a, \omega)=(1,5+1/\tau_D^2)$ and $(\sqrt{2}, 4-8/\tau_D)$
where $\tau_D(=2+\sqrt{3})$ is the characteristic ratio of these tilings.
On the other hand, the difference appears in
the weights for the third nearest-neighbor with a distance $R/a=\sqrt{3}$.
Namely, $\omega=58/\tau_D-14\sim 1.54105$ for the hexagonal tiling
while $\omega=33/\tau_D-3\sim 1.42116$ for the dodecagonal tiling.
This difference is relatively smaller than that between the Shastry-Sutherland and trellis lattices discussed above.
In addition, even for $R/a\gtrsim \sqrt{3}$, 
the locations for the peaks are almost the same and
their weights take similar values, as shown in Fig.~\ref{fig:Stampfli}(c).
This should lead to only a slight difference in their order metrics.
Figure~\ref{fig:Stampfli}(d) shows the running averages for both tilings.
When $R/a<\sqrt{3}/2$, the curves of the running average 
are identical.
Furthermore, a difference in these curves is invisible for $R/a<10$.
On the other hand, around $R/a=100$, we find that 
the running average for the dodecagonal tiling is smaller than the other.
By means of the filter function with $\sigma/a=5$, 
we obtain $\bar{\Lambda}_2(R)$ for both tilings,
which are shown as the dashed lines in the inset of Fig.~\ref{fig:Stampfli}(d).
We hardly find oscillation behavior in $\bar{\Lambda}_2(R)$ for both tilings,
and obtain the order metrics 
$\bar{B}=0.51785$ for the hexagonal tiling and
$\bar{B}=0.51764$ for the dodecagonal tiling (also see Tab.\ref{2}).
The quasiperiodic point pattern with the higher rotational symmetry has a smaller order metric,
similarly to the results for the periodic point patterns discussed in the previous section.

\begin{figure}[htb]
  \includegraphics*[width=\linewidth]{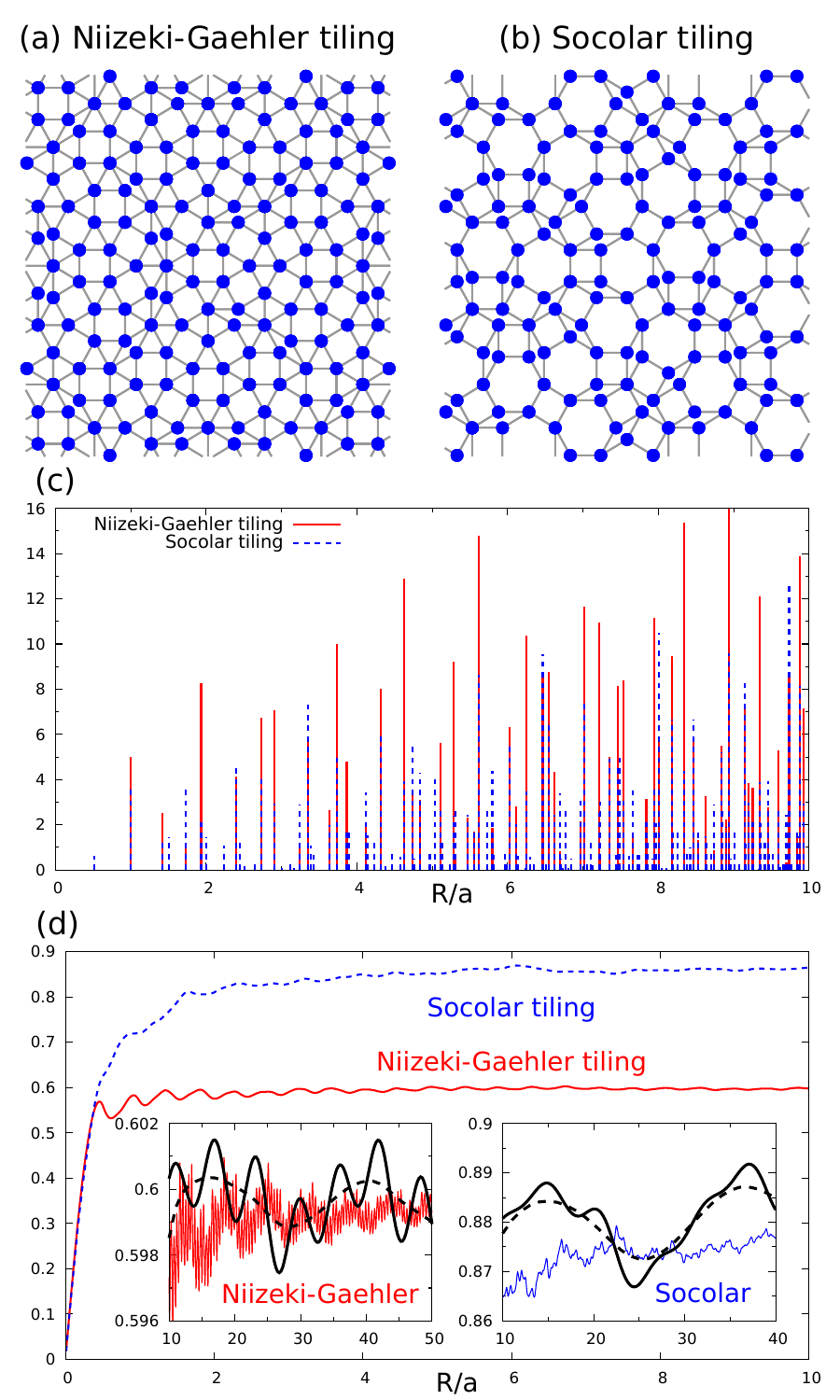}
  \caption{
    (a) Niizeki-G\"ahler and (b) Socolar tilings,
    (c) their histograms. 
    (d) Solid (dashed) lines represent the running average $\bar{\Lambda}_1(R)$
    for the Niizeki-G\"ahler (Socolar) dodecagonal tiling. 
    Left and right insets show the results for Niizeki-G\"ahler and Socolar tilings 
    in the larger $R$ case.
    Thin solid line represents the running average $\bar{\Lambda}_1(R)$, and
    Black bold solid (dashed) line represents $\bar{\Lambda}_2(R)$ with $\sigma/a=3$ ($\sigma/a=5$).
  }
  \label{fig:12}
\end{figure}

Next, we consider the Niizeki-G\"ahler~\cite{Niizeki_1987,Gahler_1988} and Socolar~\cite{Socolar} tilings,
as examples of dodecagonal tilings.
These are shown in Figs.~\ref{fig:12}(a) and \ref{fig:12}(b).
The former is composed of squares, triangles, and rhombuses.
It is known that this tiling is a key structure
for the two dimensional oxide quasicrystals
derived from $\rm BaTiO_3$ and $\rm StTiO_3$ on a Pt(111) substrate~\cite{Foerster_2013,Schenk_2017},
and structural properties have been discussed~\cite{Yamada_2022}.
The Socolar dodecagonal tiling is composed of squares, rhombuses, and hexagons,
and thereby the vertex system is bipartite.
Its magnetic properties have been discussed~\cite{Koga_2021,Keskiner_Oktel_2022}.
As shown in Figs.~\ref{fig:12}(a) and \ref{fig:12}(b),
the vertices in the Niizeki-G\"ahler tiling look densely distributed,
compared with the vertices in the Socolar tiling.
Therefore, the weights in the histogram for the Niizeki-G\"ahler tiling are 
significantly higher than the others at particular distances, as shown in Fig.~\ref{fig:12}(c).
This should affect the convergence of the running averages $\bar{\Lambda}_1(R)$.
Figure~\ref{fig:12}(d) shows the running averages for both tilings.
The running average for the Niizeki-G\"ahler tiling oscillates around $0.6$ for $R/a>2$.
By means of the filter function method with $\sigma/a=5$,
we obtain $\bar{\Lambda}_2(R)$ which are shown as the dashed lines in the insets. 
In contrast to the cases of Penrose and Stampfli tilings,
we find larger oscillations in $\bar{\Lambda}_2(R)$.
This means that the Gaussian filter with a larger $\sigma$ will be necessary to precisely evaluate
the order metric
although it is hard to obtain the histogram for large $R$
due to its large computational cost.
The less accurate order metric is obtained as $\bar{B}\sim 0.599(2)$,
by taking intro account the width of the oscillation in $\bar{\Lambda}_2(R)$.
Figure~\ref{fig:12}(d) shows that the running average for the Socolar tiling
slowly increases even when $R/a\gtrsim 5$.
In the right inset of Fig.~\ref{fig:12}(d),
a fairly large oscillation appears in $\bar{\Lambda}_2(R)$ and $\bar{\Lambda}_2(R)$ is larger than $\bar{\Lambda}_1(R)$
in most of the range $(R/a<40)$.
This should be explained by the following.
In the tiling, there is the wide space in each hexagon
and points are unevenly concentrated around rhombuses.
This short-range uneven distribution leads to the slow increase in $\bar{\Lambda}(R)$ for small $R$,
which strongly affects the running average with larger $R$.
We then obtain the less accurate order metric $\bar{B}\sim0.88(1)$,
taking into account oscillation behavior in $\bar{\Lambda}_2(R)$.

\begin{widetext}
  \begin{center}
    \begin{table}[h]
      \caption{Densities of points $\rho$, packing factor $F$, and normalized order metric $\bar{B}$ 
      for various quasiperiodic point patterns with $n$-fold rotational symmetry. 
      $\tau[=(1+\sqrt{5})/2]$ is the golden ratio, $\tau_s(=1+\sqrt{2})$ is the silver ratio,
      and $\tau_D(=2+\sqrt{3})$ is the ratio characteristic of the dodecagonal tilings.}
      \label{2}
      \centering
      \renewcommand\arraystretch{1.2}
      \setlength{\tabcolsep}{10pt} 
      \begin{tabular}{l|c|ll|ll|l}
          \toprule
          Point pattern & $n$ &$\rho$ && $F$ &&$\bar{B}$ \\
        \hline
        Square Fibonacci tiling& 4 & $\tau^2/5$ & 0.523 61 & $\pi\tau^2/20$ &  0.411 24 & 0.835(5) \\
        Hexagonal 3-tile tiling& 6 & $8\tau/(15\sqrt{3})$ & 0.498 23 & $2\pi\tau/(15\sqrt{3})$ & 0.391 31 & 1.39(3) \\
        Stampfli hexagonal tiling& 6 & $\tau_D/(2\sqrt{3})$ & 1.077 35 & $\pi\tau_D/(8\sqrt{3})$ & 0.846 15 & 0.517 85\\
        Ammann-Beenker tiling& 8 & $\tau_s/2$ & 1.207 11 & $\pi/(4\sqrt{2}) $ & 0.555 36 & 0.590(1) \\
        Penrose tiling& 10 & $2\cdot 5^{-3/4} \tau^{3/2}$ & 1.231 07 & $5^{-3/4} \pi /(2\tau^{1/2})$ & 0.369 32 & 0.591 45\\
        Stampfli dodecagonal tiling& 12 & $\tau_D/(2\sqrt{3})$ & 1.077 35 & $\pi\tau_D/(8\sqrt{3})$ & 0.846 15 & 0.517 64 \\
        Niizeki-G\"ahler tiling& 12 & $2/\sqrt{3}$ & 1.154 70 & $\pi/(2\sqrt{3}\tau_D)$ & 0.243 00 & 0.599(2) \\
        Socolar dodecagonal tiling& 12 & $2(3+\sqrt{3})/9$ & 1.051 57 & $(3-\sqrt{3})\pi/18$ & 0.221 30 & 0.88(1)\\
          \toprule
      \end{tabular}
    \end{table}
  \end{center}
\end{widetext}

Here, we have examined the hyperuniformity for three dodecagonal tilings
and have found that Stampfli, Niizeki-G\"ahler, and Socolar dodecagonal tilings
in descending order of order metric.
This result suggests that
the order metric is correlated with the packing factor rather than the density of points
since the Niizeki-G\"ahler and Socolar tilings include 
the rhombuses with acute angles of $\pi/6$ 
and their packing factors are less than that of the Stampfli tiling.
The density of points $\rho$, packing factor $F$, normalized order metric $\bar{B}$
for several quasiperiodic tilings are explicitly shown in Tab.~\ref{2}.

\begin{figure}[htb]
  \includegraphics*[width=\linewidth]{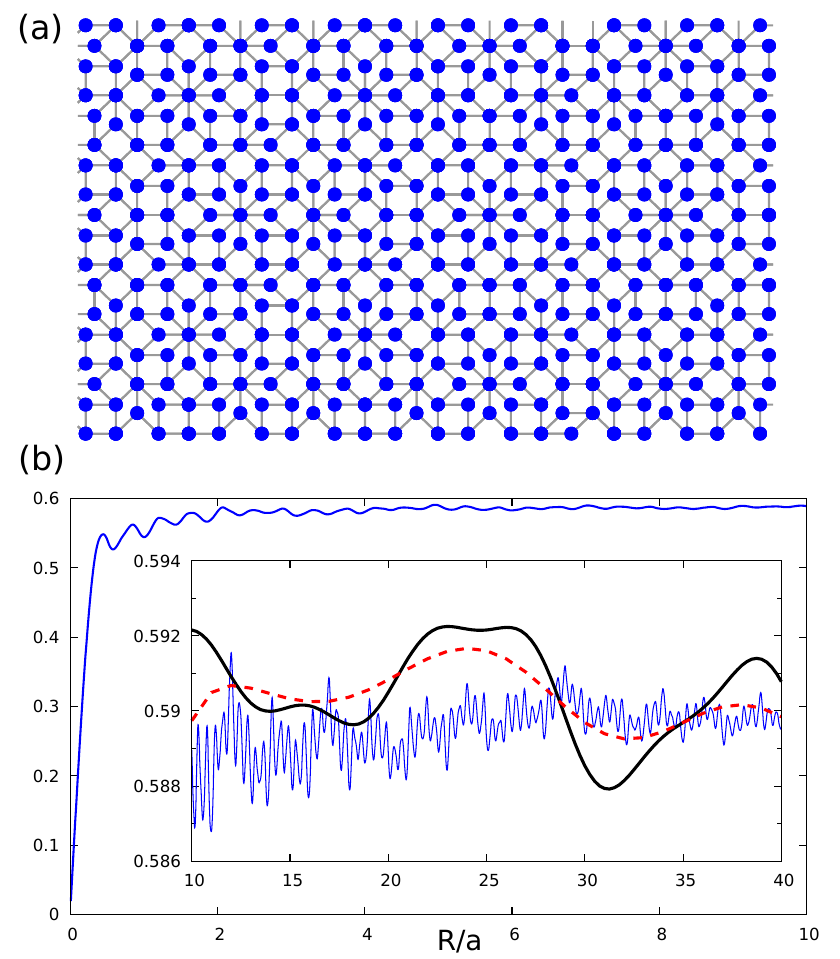}
  \caption{
    (a) Ammann-Beenker tiling.
    (b) Thin solid line represents the running average $\bar{\Lambda}_1(R)$ and
    the bold solid (dashed) line represents $\bar{\Lambda}_2(R)$ obtained 
    by the Gaussian with $\sigma/a=3$ ($\sigma/a=5$).
  }
  \label{fig:AB}
\end{figure}

We also study the Ammann-Beenker tiling~\cite{Socolar,Baake}, as shown in Fig.~\ref{fig:AB}(a).
This tiling is composed of the squares and rhombuses, 
and is invariant under eightfold rotation operations.
Figure~\ref{fig:AB}(b) shows the running average as a function of $R/a$,
where $a$ is the edge length of the tiles.
We find that the running average seems to converge for $R/a\sim 3$  
while the sampling number dependence appears in the large $R$ region (not shown), 
which is similar to that for the Niizeki-G\"ahler and Socolar tilings discussed above.
It is not so clear why the convergence of the histogram strongly depends on the tilings.
By means of the filter functions,
we obtain the order metrics 
$\bar{B}=0.590(1)$ for the Ammann-Beenker tiling,
which is smaller than that in the previous work~\cite{Zachary_2009}.

\begin{figure}[htb]
  \includegraphics*[width=\linewidth]{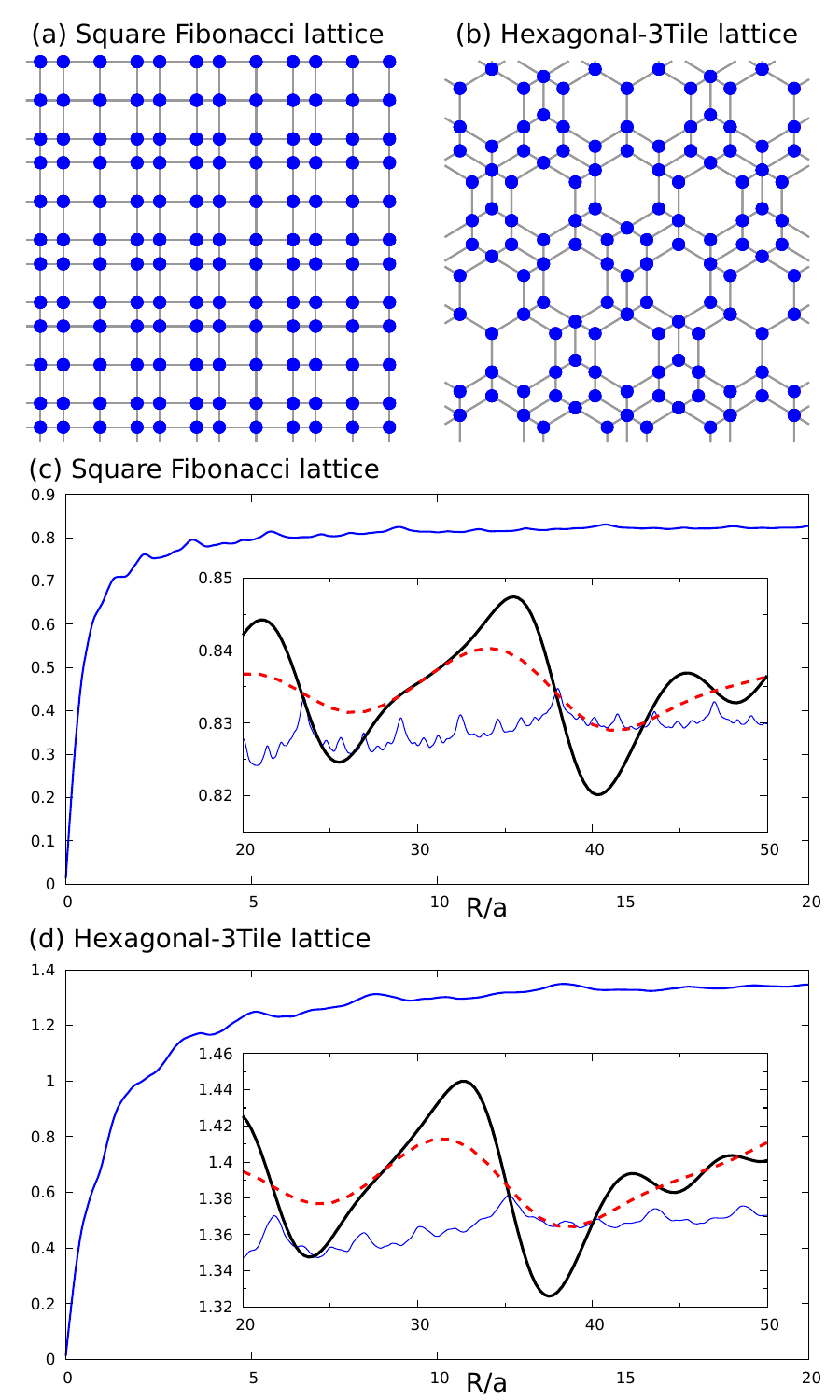}
  \caption{
  (a) Square Fibonacci and (b) hexagonal three-tile lattices.
  (c) [(d)] Blue thin curve shows the running average for the square Fibonacci 
  (hexagonal three-tile) lattice.
  Bold solid (dashed) curves in the insets of (c) and (d) represent
  $\bar{\Lambda}_2(R)$ obtained by the Gaussian with $\sigma/a=3$ ($\sigma/a=5$).
  }
  \label{fig:2len}
\end{figure}
Finally, we consider the quasiperiodic tilings with two length scales
and examine the order metrics for the point patterns.
One of the simplest tilings is the squared Fibonacci tiling~\cite{Lifshitz_2002},
where the one-dimensional Fibonacci sequences are loaded into
the edges of the square lattice in both horizontal and vertical directions,
as shown in Fig.~\ref{fig:2len}(a).
The tiling is composed of the small and large squares, and rectangles.
The ratio between short and long edges is set as the golden ratio.
Recently, the hexagonal three-tile tiling,
which is composed of small and large hexagons, and parallelograms,
has been found~\cite{Sam}.
The ratio between the short and long edges is given by the golden ratio.
Figure~\ref{fig:2len}(c)[(d)] shows the running average for
square Fibonacci (hexagonal 3-tile) tiling.
Similar to the Ammann-Beenker, Niizeki-G\"ahler, and Socolar tilings,
the running averages slowly increase, in contrast to the Penrose and Stampfli tilings.
We obtain $\bar{B}=0.835(5)$ for the square Fibonacci tiling and
$\bar{B}=1.39(3)$ for the hexagonal three-tile tiling
by means of the filter function method with $\sigma/a=5$.

\section{Summary}\label{sec:summmary}

We have studied the hyperuniformity
of the two-dimensional periodic and quasiperiodic point patterns systematically.
To calculate the hyperuniformity order metric, 
which quantifies the regularity of the hyperuniform point patterns,
we have developed an efficient method,
where the filter function and histogram of two-point distances are combined.
Then, we have calculated the hyperuniformity order metric.
For the Shastry-Sutherland and trellis lattices composed of
triangles and squares, we have demonstrated that the order metric for the former
is smaller than the latter.
We have also compared the order metrics for
the Stampfli hexagonal and dodecagonal quasiperiodic tilings.
The order metric for the former is larger than the latter.
These results indicate that the order metric is deeply related to
the lattice symmetry, 
{\it i.e.}, being smaller for a higher symmetry,
in addition to the density of points.
The filter-function method proposed here will also be applicable to density (scalar-field) distributions~\cite{Ma_Torquato_2017}:
Hyperuniform density distributions appear in electron systems on quasiperiodic lattices~\cite{Sakai_Arita_Ohtsuki_2022,Sakai_2022}.
Since the system size tractable with numerical simulations is limited, the method may be particularly useful in computing the order metric of such distributions.

\begin{acknowledgments}
  We would like to thank N. Fujita, T. Ishimasa, and T. Yamada for valuable discussions.
  Parts of the numerical calculations were performed
  in the supercomputing systems in ISSP, the University of Tokyo.
  This work was supported by Grant-in-Aid for Scientific Research from
  JSPS, KAKENHI Grant Nos. JP22K03525, JP21H01025, JP19H05821 (A.K.), and 22H04603 (S.S.).
\end{acknowledgments}



\bibliography{./refs}

\end{document}